# DRISHTI: Visual Navigation Assistant for Visually Impaired

**Malay Joshi\*, Aditi Shukla, Jayesh Srivastava, Manya Rastogi**

Department of Electronics and Communication Engineering,

Ajay Kumar Garg Engineering College, Ghaziabad, Uttar Pradesh, 201009, India

\*Corresponding author e-mail: malay1931025@akgec.ac.in

**Abstract.** In today's society, where independent living is becoming increasingly important, it can be extremely constricting for those who are blind. Blind and visually impaired (BVI) people face challenges because they need manual support to prompt information about their environment. In this work, we took our first step towards developing an affordable and high-performing eye wearable assistive device, DRISHTI, to provide visual navigation assistance for BVI people. This system comprises a camera module, ESP32 processor, Bluetooth module, smartphone and speakers. Using artificial intelligence, this system is proposed to detect and understand the nature of the users' path and obstacles ahead of the user in that path and then inform BVI users about it via audio output to enable them to acquire directions by themselves on their journey. This first step discussed in this paper involves establishing a proof-of-concept of achieving the right balance of affordability and performance by testing an initial software integration of a currency detection algorithm on a low-cost embedded arrangement. This work will lay the foundation for our upcoming works toward achieving the goal of assisting the maximum of BVI people around the globe in moving independently.

## 1. Introduction

Blindness is a daunting condition. According to a report by the WHO in 2013, there is an estimated 40 to 45 million people who are blind, and about 135 million have low or weak sight. According to a report by The Hindu, 62 million people in India are visually impaired, of which eight million are blind. Visual impairment can impact a person's quality of life and make them prone to discrimination. They face many challenges in navigating around places. There is a large number of adaptive equipment that enable visually impaired people to live their life independently. However, they are only found in nearby shops or marketplaces. Also, they are quite expensive, so only some BVI people can use such resources.

The camera-operated mechanism in [1] helps them read text on things that are held in their hands easily. The proposed system uses a camera to capture the target object, and an algorithm extracts the text in the captured image from the backdrop. Each text letter is separated by optical character recognition (OCR). Audio output is provided using a software development kit with the identified text. The wearable system in [2] is a device that receives user input and recognises things. The device has an ultrasonic sensor that assists in warning the user of objects that are in his or her path. The items are located using the Haar cascade method. It is a wearable device that can be mounted on the user's chest. The user will receive an audio of the object that was discovered.

The traffic scenes use object detection technology in [3] to find objects. Here, they have used a combination of R-FCN (Regression-based Full Convolution Network) and OYOLO (optimized you only look once), which is 1.18 times faster than YOLO. It identifies and categorizes photos of vehicles, cyclists, and other objects. Location errors occur when using YOLO; to prevent these, we employ OYOLO. Other possible categories of solutions analyzed in [4], with promising present and future utilities, are based on the existing technologies.

The system in [5] uses features like Artificial Intelligence and Machine Learning to provide a solution to the problem. A device camera captures images, objects are detected, and distance is calculated using an application. The prototype in [6] is mounted on top of a walking cane that uses a pi cam to click pictures and then implements the YOLO algorithm to perform object detection that works more accurately than others. After that, the gTTS module converts text to speech, resulting in a human voice.

All currently available systems have one or more of the following drawbacks: (i) unaffordable price (ii) lack of sales, marketing, or servicing in developing nations; (iii) high inaccuracy, thus making them unfit for general usage; and/or (iv) bulky or challenging to use.

There are five sections in the paper. Section 1 covers the introduction to the paper and some existing works in the domain of virtual assistance devices. Section 2 presents survey results done as part of the idea validation process. Section 3 contains the hardware and software details of the proposed system, and section 4 covers the real-time test results of the proposed system. The paper is concluded in section 5. Section 6 provides the future scope of this project. The references are there in section 7.

## 2. In-person survey analysis
After surveying BVI people at a local trust for blind people, the results obtained were:
(1) They all had no impairment other than visual impairment.
(2) There are apps for currency detection, colour recognition and document reading.
(3) There is also a read-aloud feature in their smartphones named Talkback that helps them to use the smartphones.
(4) Their current device is a cane with an ultrasonic sensor that detects obstacles, but the limitation is that it does not give the details of whether it is a dog or a heap of stones.
(5) Their cane is based on touch, and its processing is slow.
(6) The cane costs around INR 4000, and sometimes they do not run after branded items.
(7) They all agreed to wear spectacles weighing up to 100 grams.
(8) Mostly everyone has a smartphone and a good internet connection.
(9) According to them, if a device helps 60% of visually challenged people, it is considered useful and successful.

## 3. Proposed system
This is a microcontroller-based virtual visual assistance device for visually challenged people. The main objective of this system is to convert the visual world into an audio world to make it easier for people with visual disabilities to navigate themselves and do their daily activities without feeling deprived of vision.

The ESP32 camera module will capture images of the target(s) (like surrounding objects, people, text document, currency, road, traffic signal, traffic sign board, etc). This captured image will be sent to a smartphone device and processed in real-time by multiple algorithms. The information extracted will be converted into audio-based signals that will act as user feedback.

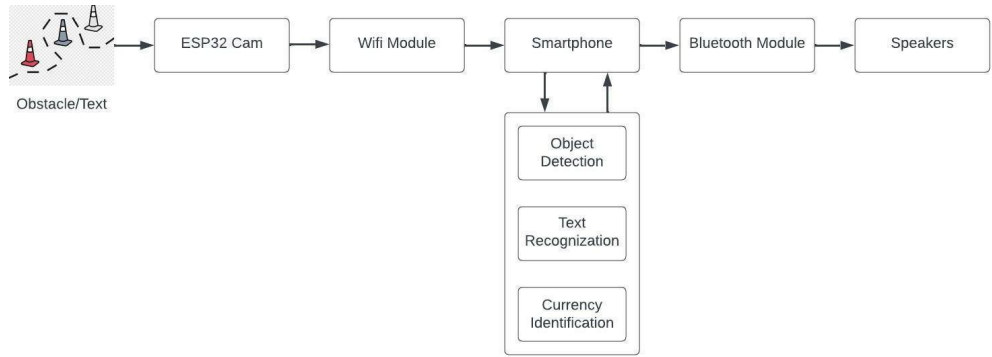

**Fig 1**. System Block Diagram

*3.1. Hardware used*
Hardware components used in this proposed solution are:
(1) ESP32 Cam: It is an ESP32-based development board that is low-cost and comes with a small onboard camera. It is an ideal solution for small IoT applications, constructions, prototypes, and DIY projects. The camera module board integrates features like traditional Bluetooth and low power BLE, WiFi, and is equipped with two high-performance 32-bit LX6 CPUs.
(2) Programmer Module: The ESP32-CAM has an OV2640 camera, onboard flash, microSD card support option and several GPIOs that can connect peripherals. However, it does not have a built-in programmer. So it needs an external programmer to connect it to the computer and upload the code. A Future Technology Devices International (abbreviated as FTDI) programmer or any Arduino board can be used to implement this task. It directs the camera sensor to capture images and send them to the desired location for processing.
(3) Smartphone: The sent images are received by any smartphone device, and different algorithms for object detection, text recognition and currency identification are implemented. For these purposes, machine learning and artificial intelligence are needed.
(4) Speakers: After processing the image, the output is generated in the form of voiced instructions. This is done with the help of two 3Watt Wireless Bluetooth Multimedia Speakers connected using an HC-05 Bluetooth module. This whole process is done in real-time, and the voiced instructions will help a blind person in navigation and text reading.

*3.2 Software used*
Algorithms and modules used in this proposed solution are as follows:
(1) Deep Learning (Resnet-50): It is used for detection and identification of currency. Resnet-50 is a 50 layer deep convolutional neural network. ResNet stands for Residual Network. It solves complex problems with improved accuracy and performance. Basically, deep learning is a way to enhance human gain like used for automated driving, stop signs ,and traffic lights.
(2) YOLOv7: You Only Look Once (YOLO) is a common algorithm for object detection. It is famous for detecting objects in a real time environment. Used for detecting traffic signals, exam proctoring etc. The best model of YOLOv7 scored 56.8% Average Precision (AP), which according to the paper is the highest among all known object detectors.
(3) Tesseract OCR: Tesseract is a printed text reader. It is an engine for optical character recognition used by various OS. OCR creates a new searchable text file or a PDF by extracting text from images and documents without a text layer.. Tesseract has very improved image quality.

(4) gTTS: Google Text-to-Speech (gTTS) is used for speech translation. The text-to-speech API of Google Translate is interfaced with using a Python library and CLI tool. The text variable is a string used to store the user's input. Languages including English, Hindi, Tamil, French, German, and many more are supported via the gTTS API.

## 4. Results
*4.1. Circuit diagram*

An FTDI programmer is used to program the ESP32 Cam. In this circuit, ESP32 Cam module is connected to FTDI module by connecting pins GND, U0T, U0R of ESP32 Cam module with GND, RX and TX pins of FTDI module respectively. This whole setup is powered by a 5V battery. Note: GPIO 0 pin of the esp32 cam module must be connected to GND when uploading code.

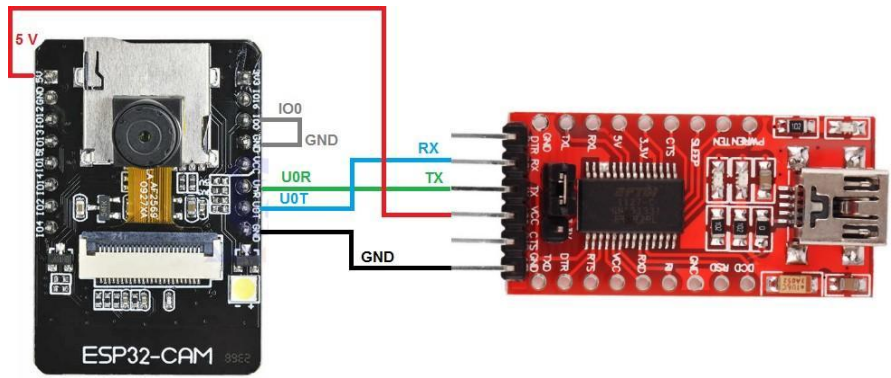

**Fig 2.** Connecting ESP32Cam with Future Technology Devices International (FTDI) module [7]

*4.2. Software Simulation*

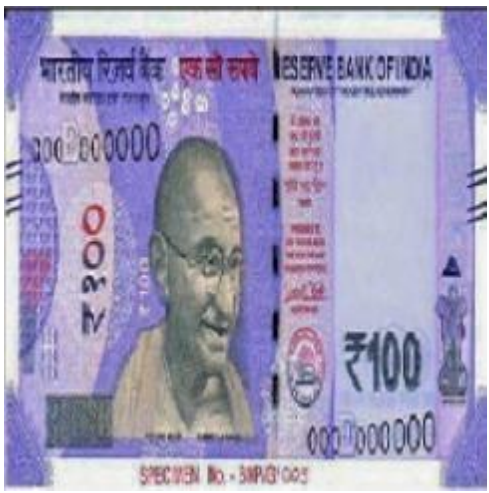

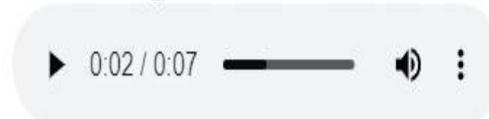

**Fig 3.** Image of INR 100 given as input       **Fig 4.** Model's text and audio-based prediction

(1) Case-I. When we place a note of Rs 100, it acts as an input. It captures the image, processes it and then applies the algorithm. It detects the denomination and gives the output as 100 with a probability of 1. The output received is both in the form of audio and text.

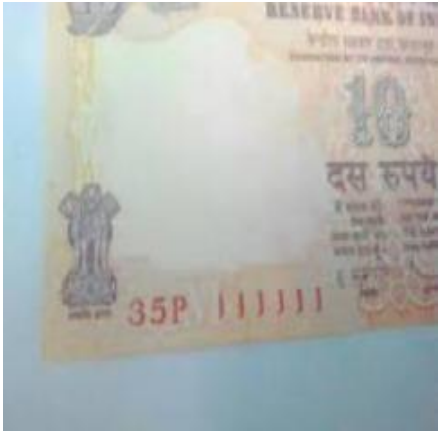
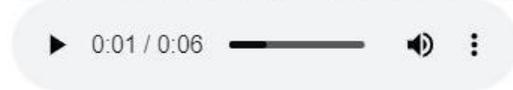

**Fig 5.** Image of INR 10 given as input    **Fig 6.** Model's text and audio-based prediction

(2) Case-II. When a 10 rupees note is given as an input, it captures the image and analyses it. After processing the image, the algorithm is applied. It gives output as 10 with a likelihood or probability of 0.89. The output is given in the form of audio as well as text.

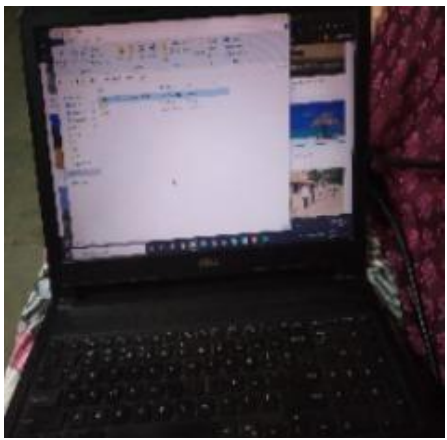
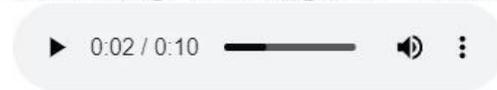

**Fig 7**. Image of laptop given as input    **Fig 8.** Model's text and audio-based prediction

(3) Case-III. When the above background image is given as input, it captures the image and processes it. After image processing, it applies an algorithm. It produces output that is a background with a likelihood of 1. The output is given in the form of audio as well as text.

**5. Conclusion**
Due to the lack of directly perceiving visual information, many blind and visually impaired people struggle to maintain a constant healthy rhythm. Therefore, a navigation system that enables blind people to navigate their route freely and lets them know where they might be at any given time is necessary. To make the life of visually impaired people easier, we have used technology in this article to give them a visual aid. The project's main goal is to design an object detector that can recognise obstructions and guide a visually impaired person in a path via voiced instructions. Hence, the design and implementation of a cost-effective device were done to provide support and independence to visually impaired people during their travel to new or unknown places.

## 6. Future scope

Future work will include improvements to the device's design to make it affordable for commercial use, adding computer vision-based algorithms for analysing the nature of travel paths and obstacles ahead, and conducting user research to improve the system's usability as a whole. These future works will help blind and visually challenged people in independent navigation.